\documentclass[12pt]{article}
\usepackage{times}
\usepackage{geometry}
\usepackage[utf8]{inputenc}
\usepackage{enumitem,amssymb}
\usepackage{ragged2e}
\usepackage{xcolor,wasysym,natbib,hyperref,graphicx}
\newlist{thematic}{itemize}{8}
\setlist[thematic]{label=$\square$}
\usepackage{pifont}

\def\aj{{AJ}}                   
\def\araa{{ARA\&A}}             
\def\apj{{ApJ}}                 
\def\apjl{{ApJ}}                
\def\apjs{{ApJS}}               
\def\ao{{Appl.~Opt.}}           
\def\aap{{A\&A}}                

\def\mnras{{MNRAS}}             
\def\pasa{{PASA}}               
\def\nat{{Nature}}              

\begin{document}
\raggedright
\huge
Astro2020 Science White Paper \linebreak

In Pursuit of Galactic Archaeology \linebreak
\normalsize

\noindent \textbf{Thematic Areas:} \hspace*{60pt} $\square$ Planetary Systems \hspace*{10pt} $\square$ Star and Planet Formation \hspace*{20pt}\linebreak
$\square$  Formation and Evolution of Compact Objects \hspace*{31pt} $\square$ Cosmology and Fundamental Physics \linebreak
  $\XBox$  Stars and Stellar Evolution \hspace*{1pt} $\square$ Resolved Stellar Populations and their Environments \hspace*{40pt} \linebreak
  $\XBox$    Galaxy Evolution   \hspace*{45pt} $\square$             Multi-Messenger Astronomy and Astrophysics \hspace*{65pt} \linebreak
  
\textbf{Principal Author: Melissa Ness}

Name:	Melissa Ness
 \linebreak						
Institution:  Columbia/Flatiron
 \linebreak
Email: melissa.ness@columbia.edu
 \linebreak
 
\textbf{Co-authors:} 
  \linebreak
  Jonathan Bird (Vanderbilt),
   Jennifer Johnson (Ohio State University),
  Gail Zasowski (University of Utah),
  Juna Kollmeier (Carnegie),
  Hans-Walter Rix (MPIA),
  Victor Silva Aguirre (Aarhus),
   Borja Anguiano (University of Virginia),
   Sarbani Basu (Yale),
      Anthony Brown (Leiden),
Sven Buder (MPIA),
  Cristina Chiappini (AIP),
    Katia Cunha (NOAO),
  Elena Dongia (University of Wisconsin, Madison),
  Peter Frinchaboy (TCU),
Saskia Hekker (MPI for Solar system research),
  Jason Hunt (Toronto),
  Kathryn Johnston (Columbia) ,
    Richard Lane (PUC),
    Sara Lucatello (INAF Osservatorio Astronomico di Padova),
    Szabolcs Meszaros (ELTE Gothard Astrophysical Observatory)
      Andres Meza (UDD),
  Ivan Minchev (AIP),
  David Nataf (JHU),
  Marc Pinsonneault (Ohio State University),
      Adrian M. Price-Whelan (Princeton),
      Robyn Sanderson (UPenn/Flatiron)
    Jennifer Sobeck (University of Washington),
  Keivan Stassun (Vanderbilt),
  Matthias Steinmetz (AIP),
Yuan-Sen Ting (IAS/Princeton/OCIW),
  Kim Venn (Victoria),
  Xiangxiang Xue (NAOC)
 \linebreak


\textbf{Abstract: 
The next decade affords tremendous opportunity to achieve the goals of Galactic archaeology. That is, to reconstruct the evolutionary narrative of the Milky Way, based on the empirical data that describes its current morphological, dynamical, temporal and chemical structures. Here, we describe a path to achieving this goal.  The critical observational objective is a Galaxy-scale, contiguous, comprehensive mapping of the disk's phase space, tracing where the majority of the stellar mass resides. An ensemble of recent, ongoing, and imminent surveys are working to deliver such a transformative stellar map. Once this empirical description of the dust-obscured disk is assembled, we will no longer be operationally limited by the observational data.  The primary and significant challenge within stellar astronomy and Galactic archaeology will then be in {\it fully utilizing these data}. We outline the next-decade framework for obtaining  and then realizing the potential of the data to chart the Galactic disk via its stars.  One way to support the investment in the massive data assemblage will be to establish a Galactic Archaeology Consortium across the ensemble of stellar missions. This would reflect a long-term commitment to build and support a network of personnel in a dedicated effort to aggregate, engineer, and transform stellar measurements into a comprehensive perspective of our Galaxy.  
}

\pagebreak

\section{Introduction}

We are entering an era where the current difficulties in building an understanding the formation and evolution of our Galaxy can be overcome. If we can trace back the history of the Milky Way, to describe and constrain the key processes that have been relevant in its evolution over time, we will understand spiral galaxies in general in our Universe. An expansive exploration of the Milky Way is therefore relevant within a cosmological context, and the archaeological pursuit reaches far beyond the characterization of a Galaxy in isolation. There is tremendous opportunity to make giant leaps forward in the domain of Galactic archaeology of the Milky Way in the coming decade, given key investments, using stars as tools.
\vspace{4pt}

The spectroscopic surveys of the current decade, including APOGEE \citep{Majewski2017}, GALAH \citep{deSilva2015}, Gaia-ESO \citep{Gilmore2012}, RAVE \citep{Steinmetz2006}, and LAMOST \citep{Newberg2012} have been revolutionary. We have gained broad insight into the characteristics of our Galaxy and some of the processes that are relevant in its evolution \citep[see][for a review]{BH2016}. Current data has driven major breakthroughs in how we make chemical abundance and age measurements from stars in the regime of large data, which can further improve our physical models \citep[e.g.][]{Ness2015, Casey2016, Ho2017a, Ho2017b, TingPayne2018, Leung2019a}. Key to these advances has been overlap in surveys and the precision age determinations from asteroseismology \citep[e.g.][]{ Kepler2010, SilvaAguirre2012, Corot2016,  Anders2017, Miglio2017, R2017, Pins2018, SA2018}. We have also concurrently overcome challenges in combining data from different surveys consistently, with methodological advances. However, due to the sparse and limited sampling and coverage of the current generation data, we are, as of 2019, currently fundamentally limited in our ability to carry out archaeological expeditions on the Milky Way.
\vspace{4pt}

Progress in the domain of Galactic archaeology is premised upon the expectation of next generation survey data to deliver a complete stellar mapping across the disk, where the majority of the mass resides. Complementary to this, a large-scale survey of the chemistry of the local stellar halo offers the prospect of understanding the nature of stars {\it not} made in our Galaxy and the first building blocks in the early Universe.  From this data, we can examine the Galaxy in detail across a range of temporal, dynamical and spatial scales. In practice, we must build an ensemble of measurements, of ages, orbits, distances and chemical abundances, which represent the best set of numbers we can hope to derive.

\vspace{4pt} 

\section{The Empirical Map of the Milky Way}

A complete and detailed mapping of the meso-structure of the disk, across an unprecedented dynamic range in scale, with an unbroken view of the Milky Way is core to the pursuit of Galactic archaeology.  The revolutionary {\it Gaia} mission has shown that the Galactic disk displays rich signatures of dynamically driven processes on local and global scales \citep[e.g.][]{Antoja2018, Trick2018, BH2019, Khanna2019,  Ted2019b}. We currently, however, have little insight as to the structure in chemistry, age and orbits over an expansive range in scale of the Galaxy. By making an empirical map of the dust-obscured disk where the majority of the stellar mass resides we will reveal the Galaxy and its structure as never before \citep[][]{Kollmeier2017} -- see Figure 1. This data will not only constrain the processes that have forged the disk as we see it today, but also to link to other galaxies.
\vspace{4pt}

 \begin{figure}[ht!]
 \vspace{-10pt} 
 \includegraphics[width=0.5\textwidth]{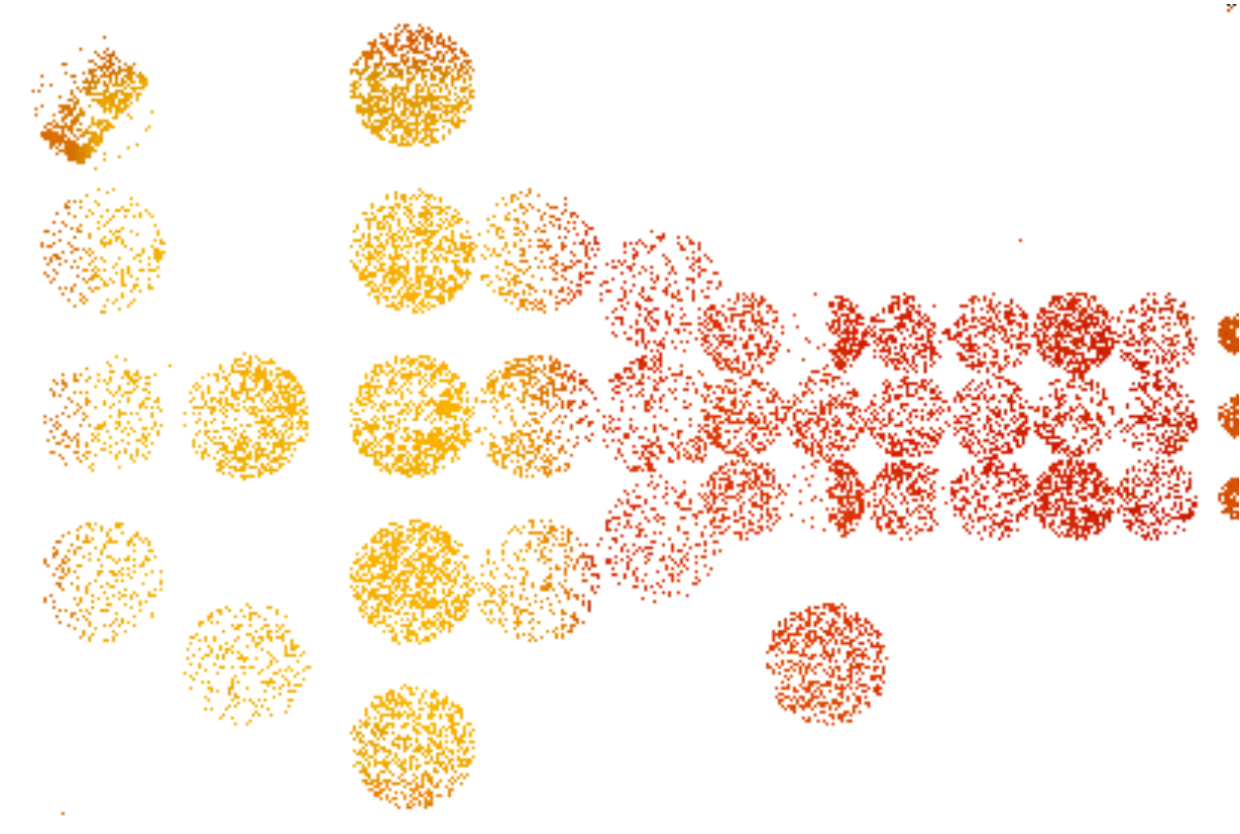}
\includegraphics[width=0.5\textwidth]{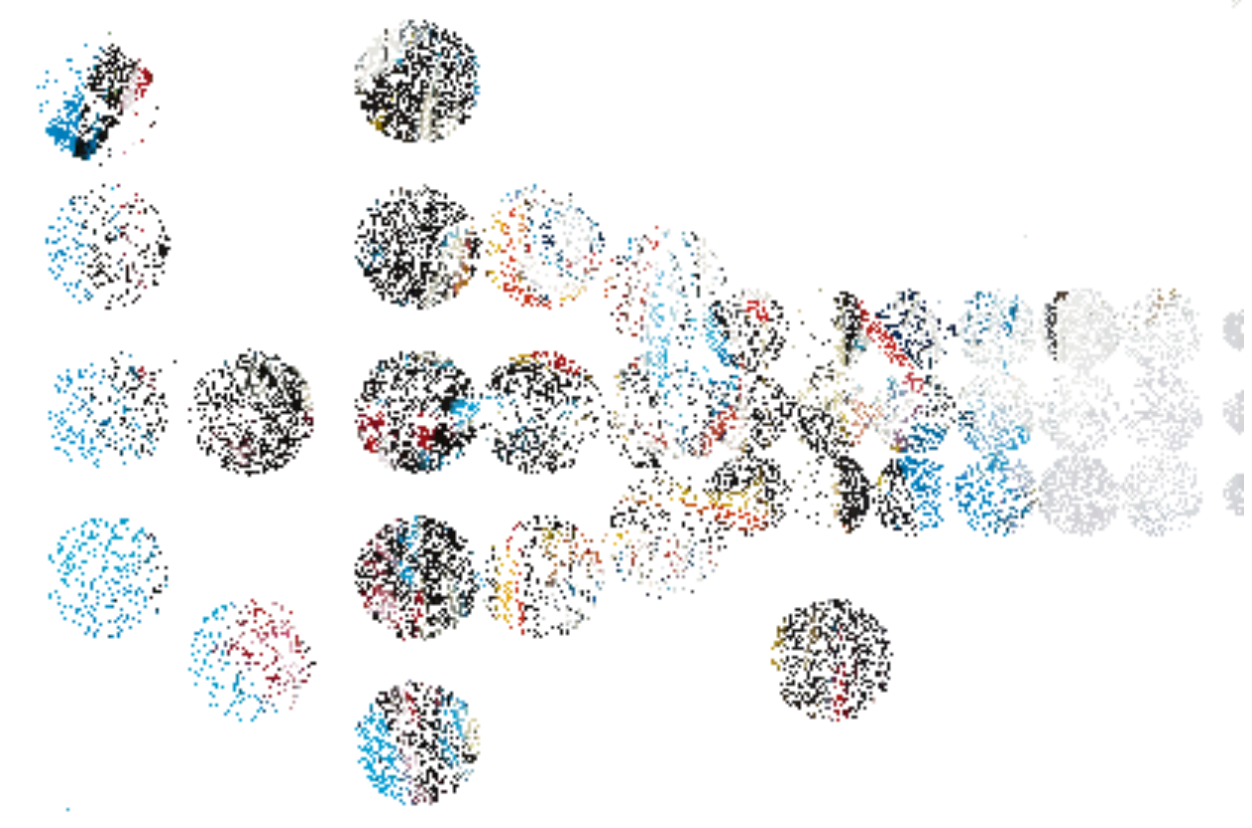}
\includegraphics[width=0.5\textwidth]{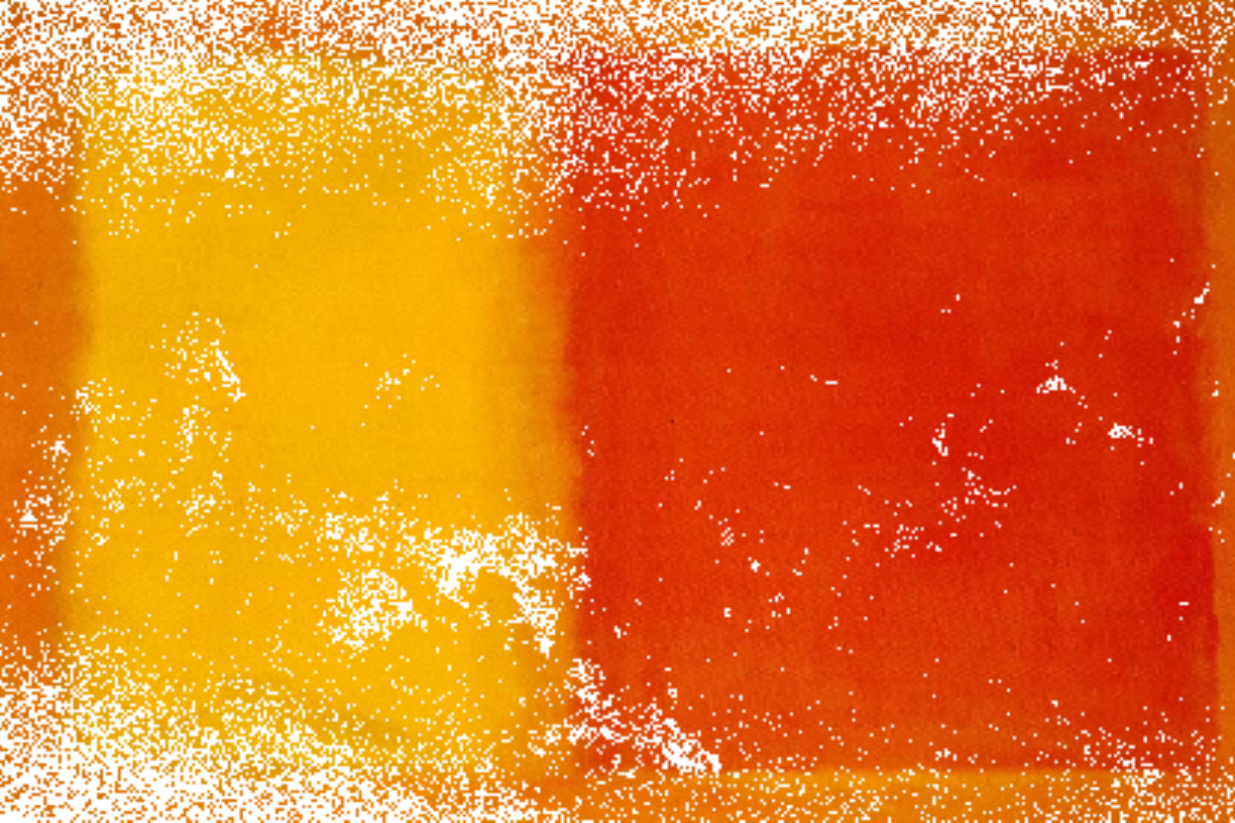}
\includegraphics[width=0.5\textwidth]{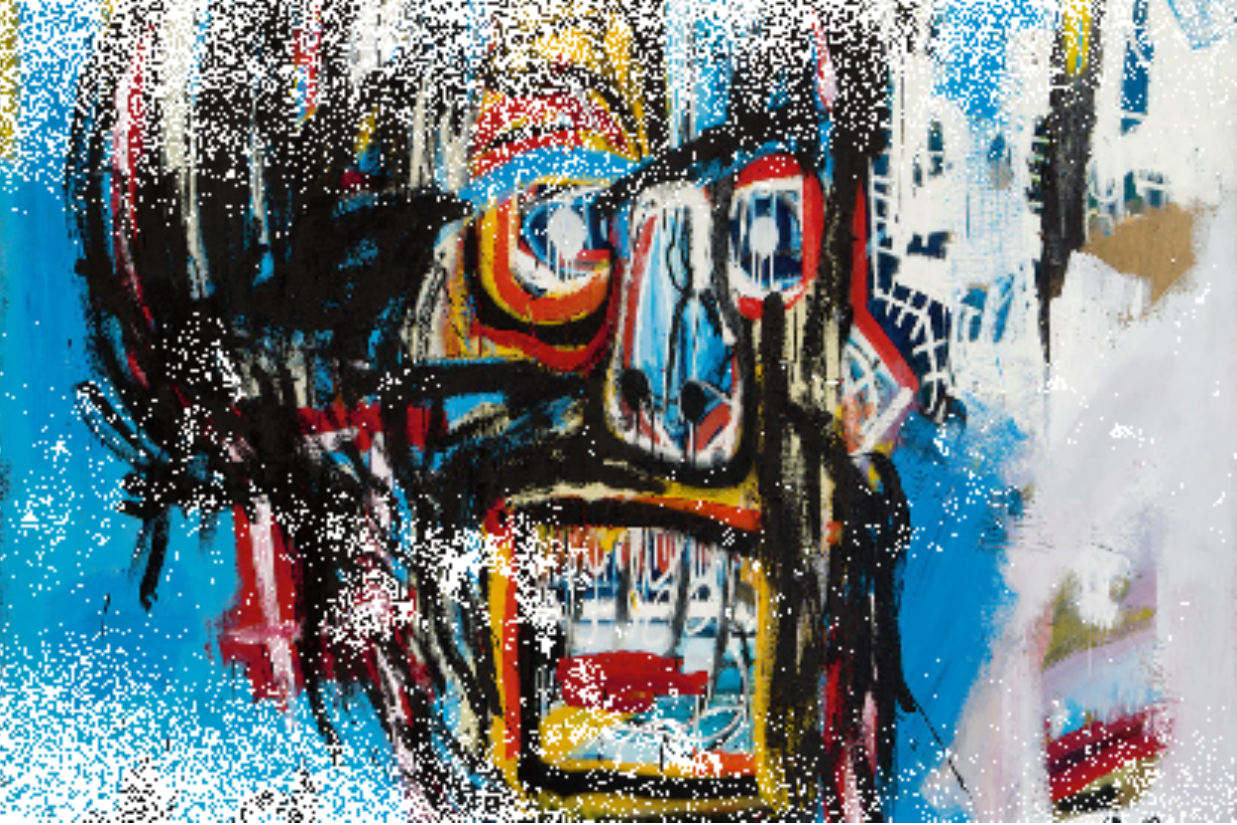}
 \caption{\footnotesize { The transition from the top to bottom panels represents the transformative view of the Galaxy that will be provided by the contiguous, continuous mapping of Sloan V's Milky Way Mapper. At top is the (pencil-beam) sampling of the groundbreaking APOGEE survey, for paintings by Mark Rothko at left and Jean-Michel Basquiat, at right. If the Milky Way is structured on the scales of Basquiat (as Gaia results indicate) then our dense mapping that will be carried out in the 2020's will revolutionize our picture of the Galaxy. If our Galaxy is sparsely structured, the gains in dense sampling will be more modest. (In practice, vast samples and many more dimensions of information get to a level of detail that even this striking illustration only partly indicates). Image sampling credit: Jonathan Bird}}
 \vspace{-5pt} 
\end{figure}

An ensemble of missions that are planned for the coming decade will build the map of our Galaxy. This includes { \it Sloan V's Milky Way Mapper} \citep{Kollmeier2017}, {\it WEAVE} \citep{Bonifacio2016}, {\it 4MOST} \citep{deJong2019}, {\it GALAH} \citep{deSilva2015}, {\it PFS} \citep{PFS2018}, {\it LAMOST} \citep{Newberg2012} and {\it MOONS} \citep{C2014}. From these data comprising many millions of stars, we can build a set of abundance measurements and ages, across a vast expanse of our Galaxy. Given new data-driven modeling approaches we also expect to deliver precision distances in the regime where {\it Gaia}  can not, enabling precise orbits to be determined across the expanse of the disk \citep{Eilers2019, Leung2019b}. As revolutionary as the results of the {\it Gaia} mission are, for most stars beyond the solar neighborhood, {\it Gaia} only provides information in limited dimensions. The spectroscopic survey complements to Gaia will greatly enhance the results of {\it Gaia} with radial velocities, chemistry, and distances, most critically, for dust-obscured stars in the inner Milky Way, where most stars live.  
 \vspace{4pt}

 This stellar census, if executed successfully and when complete, will provide the means to address numerous long-standing questions. Specifically, (1) What is the chemical and dynamical structure of the disk from its inner to outer extent?  We can answer this from what will be the highest dimensional stellar map ever realized ($>$20 chemical element abundances,  ages, 3D velocities, distances).  When combined with precise positional and photometric information from the {\it Gaia} and {\it TESS} and {\it Kepler} (and in the future {\it PLATO}) satellites, the spectra  will also enable the study of stellar physics through fundamental measurements of masses, radii, and ages.  (2) What are the distinct formation and evolutionary histories of our Galaxy and the roles of hierarchical accretion and radial migration \citep[e.g.][]{Roskar2008, Minchev2013, VC2014}? We will be able to determine birth sites and infer which evolutionary processes drove stars  from their birthplaces \citep[e.g.][]{Minchev2018, Frankel2018}. (3) What is the detailed morphological characteristic of the interstellar medium? By tracing stellar density, we examine the material between the stars, accessing the evolution and lifecycles of star forming regions across orders of magnitude range in scale. (4) Where does the Milky Way fit into detailed cosmological context? This extra-galactic view of our Galaxy can only be realized by an expansive spatial volume and  extent of a contiguous, continuous mapping, not the coarse sampling approach of previous generations of surveys.
\vspace{4pt}

\section{The Galactic Archaeology Consortium}

A significant challenge in the coming decade is managing, engineering, extracting and examining information (including of ages, masses, abundances, orbits, distances and evolutionary states) from the high dimensional multitude of stellar data across the Galaxy. A significant fiscal investment is required, to support and promote a directed community effort in working with these data. This is an investment in both personnel and a scientific vision. One way to accomplish this would be with a stellar Galactic archaeology consortium, which would seek to do highest justice to the data that have been so precisely and deliberately collected, between and across surveys. The role of this consortium would be both practical and intellectual, seeking to achieve the following: 
\vspace{4pt}

(1)  Create and maintain a database, to aggregate and centralize the ensemble of data that are available for every star. Many stars may have a multitude of (multi-epoch) spectra from different surveys, (multi-epoch) photometry and associated data products (measurements), produced by different groups. This database would be a vital resource for the wider community.
\vspace{2pt}

(2) Develop and support algorithms to deliver consistent measurements from the data across surveys on a common scale, uniform over the whole range of atmospheric parameters, chemical abundances and ages. There are existing methodologies that show this is possible, using data-driven label transfer \citep[e.g.][]{Ness2015, Ting2017}. Given significant survey overlap, this should be possible across all missions. All developed code would be made publicly available and documented. 
\vspace{2pt}

(3) Characterize systematic artifacts, which are ubiquitous characteristics of large data sets, often induced by mechanical asymmetries. Nevertheless they can be modeled and corrected for. An example of a fatal systematic for using abundances in concert is systematic abundance amplitude changes with spectrograph fiber number \citep[e.g.][]{Ness2018}. This is a consequence of a varying line spread function across the detector. Modeling systematics to best engineer the data couples to the challenge of differentiating between systematic abundance changes due to mechanical artifacts or stellar model inaccuracies, versus true astrophysical diffusion processes \citep[e.g.][]{Liu2019, Souto2019}.
 \vspace{2pt}

(4)  Do justice to the data that will be assembled, by working to quantify the extent of information captured within it. One axes of this is abundance measurements -- what do they tell us and what precision should surveys aim for in the pursuit of Galactic archaeology, and why? It has already been established from small stellar samples that individual abundances in the low-$\alpha$ stars of the disk largely indicate overall metallicity and age \citep{Bedell2018}. There is very little residual information in abundances that is not explained by these two dimensions. Precisions of $<$ 0.03 dex (and often lower) are required in order to get at this residual abundance dispersion. This dispersion represents the potential to reconstruct individual birth sites of stars given abundances alone \citep{BH2010}. Such precision across a multitude of elements, and particularly across any span of parameter space is an unreasonable objective for large stellar surveys. Instead, in the regime of large data sets of many millions of densely sampled stars, we have tremendous opportunity in working out how to optimally combine stars to reveal the characteristics of the chemodynamical structure across a range of scales \citep[e.g.][]{Harshil2019, Xiang2018}. In this pursuit, individual abundances, at a more relaxed measurement precision for individual stars ($<$ 0.1 dex), will be extremely important on a population basis, where precision mean abundances can be determined trivially with large samples, pivoting on age, overall metallicity and spatial or dynamical extent.
\vspace{2pt}

(5) Work across disciplines. Stellar astrophysics has now entered a regime where tremendous gains can be attained by building a group that traverses mathematical, engineering, scientific, data science and data inference. A successful consortium would be a diverse collective of researchers with complementary backgrounds, expertise and vision. 

\section{Conclusion}

The Milky Way provides us with a unique opportunity to cast galaxy formation in terms of individual stars -- a factor of 1 trillion in dynamic range. The ambitious surveys coming online aim to observe millions of Milky Way stars with contiguous, continuous coverage, peering through the dust to large Galactic distances. This will systematically map the entire hierarchy of structure in the Milky Way and will fully realize the scientific potential of the Milky Way as a galaxy model organism. Community effort is a primary basis for progress in this realm. This is seen most clearly in the  establishment and execution of a multitude of surveys of the Milky Way, driven by survey teams. From current data we have seen transformative gains in our understanding of our Galaxy in the past few years alone. The next decade will, if missions are carried out as planned, see the delivery of the data that will enable us to truly \textit{see} the Galaxy for the first time, in its spatial, dynamical, chemical and temporal detail. The next difficult and necessary investment will be supporting a directed effort in the community to pursue archaeology of our Galaxy. Here we propose a consortium that broadly seeks to serve all data across all stellar surveys.  We have described some of the practical aspects of such a program. Our proposal is chiefly and effectively an investment in people in the field and an endorsement of the conglomerate of efforts so far in seeking answers the most basic and fundamental questions of our Galactic origins. 

\pagebreak

\bibliography{mknbib.bib}

\end{document}